\documentclass[preprint]{aastex}

\def\kms {{\,{\rm km}\,{\rm s}^{-1}}}

\def\etal {et~al.~}

\newcommand{\FP}{\mbox {Fokker-Planck }}
\newcommand{\nbody}{\mbox {$N$-body}}
\newcommand{\msun}{\mbox {$M_\odot$}}
\newcommand{\nuesc}{\mbox {$\nu_{\rm esc}$}}

\begin{document}

\title{Monte Carlo Simulations of Globular Cluster Evolution - II. \\
Mass Spectra, Stellar Evolution and Lifetimes in the Galaxy.}

\author{Kriten J.~Joshi\altaffilmark{1}, Cody P. Nave\altaffilmark{2},
and Frederic A.~Rasio\altaffilmark{3}}

\affil{Department of Physics, Massachusetts Institute of Technology}

\altaffiltext{1}{Current address: 1 Union Square South \#17-B, 
New York, NY 10003; 
email: kjoshi@alum.mit.edu}
\altaffiltext{2}{6-201 MIT, 77 Massachusetts Ave, Cambridge, MA 02139;
email: cpnave@mit.edu.}
\altaffiltext{3}{6-201 MIT, 77 Massachusetts Ave, Cambridge, MA 02139;
email: rasio@mit.edu.}

\begin{abstract}
We study the dynamical
evolution of globular clusters using our new 2-D Monte Carlo code,
and we calculate the lifetimes of clusters in the Galactic environment.
We include the effects of a mass spectrum, mass loss in the Galactic tidal
field, and stellar evolution. We consider initial King models containing
$N=10^5 - 3\times 10^5$ stars, with the
dimensionless central potential $W_0=\,$ 1, 3, and~7, and with
power-law mass functions $m^{-\alpha}$, with $\alpha=\,$1.5, 2.5, and~3.5.
The evolution is followed up to core collapse, or disruption, whichever
occurs first.

We compare our results with those from similar calculations
using Fokker-Planck methods.
The disruption and core-collapse times of our models are
significantly longer than those of 1-D \FP models. This is
consistent with recent comparisons with direct $N$-body simulations,
which have
also shown that the 1-D \FP models can significantly overestimate the
escape rate from tidally truncated clusters. However,
we find that our results are in very good agreement with recent 2-D
\FP calculations, for a wide range
of initial conditions, although our Monte Carlo models have a slightly
lower mass loss rate.
We find even closer agreement of our results with modified \FP calculations
that take into account the finite nature of the system.

In agreement with previous studies,
our results show that the direct mass loss due to stellar evolution
can significantly accelerate the mass loss rate through the tidal
boundary, by reducing the binding energy of the cluster and making
it expand.
This effect causes most clusters with a low initial central concentration
($W_0 \la 3$) to disrupt quickly in the Galactic tidal field.
The disruption is
particularly rapid in clusters with a relatively flat mass spectrum.
Only clusters born with high central concentrations ($W_0 \ga 7$),
or with very steep initial mass functions ($\alpha \ga 3.5$)
are likely to survive to the present and undergo core collapse.
We identify the mechanism by which clusters disrupt as a dynamical
instability in which the rate of mass loss increases catastrophically
as the tidal boundary moves inward on the crossing timescale.

To understand the various processes that lead to the escape of stars,
we study the velocity distribution and orbital characteristics of
escaping stars. We also compute the lifetime
of a cluster on an eccentric orbit in the Galaxy, such that it
fills its Roche lobe only at perigalacticon. We find that such an
orbit can extend the disruption time by at most a factor of a few
compared to a circular orbit in which the cluster fills its Roche lobe
at all times.
\end{abstract}


\section{Introduction}

The development of numerical methods for simulating the dynamical
evolution of dense star clusters in phase space
started in the 1970's with Monte Carlo techniques
(Henon 1971a,b; Spitzer 1987, and references therein),
and several groups applied these techniques
to address problems related to the evolution of globular clusters.
A method based on the direct numerical integration of the \FP\
equation in phase space was later developed by Cohn (1979, 1980).
The \FP (hereafter F-P) methods have since been greatly improved, and
they have been extended to more realistic simulations that take into
account (approximately) the presence of a mass
spectrum and tidal boundaries (Takahashi 1995, 1996, 1997;
Takahashi \& Portegies Zwart 1998, 1999), binary interactions
(Gao \etal 1991; Drukier \etal 1999), gravitational shock heating
by the galactic disk and bulge (Gnedin, Lee, \& Ostriker 1999),
and mass loss due to stellar evolution
(see Meylan \& Heggie 1997 for a recent review).
Direct $N$-body simulations can also be used to study globular cluster
dynamics (see Aarseth 1999 for a recent review),
but, until recently, they have been limited to rather unrealistic
systems containing very low numbers of stars. The GRAPE family of
special-purpose computers now make it possible to perform direct
$N$-body integrations for clusters containing up to $N\sim 32,000$
single stars, although the computing time for such large simulations
remains considerable (see Makino \etal 1997, and references therein).
This is the second of a series of papers in which we study globular
cluster dynamics using a Monte Carlo technique similar to the original
Henon (1971b) method.
Parallel supercomputers now make it possible for the first time
to perform Monte Carlo
simulations for the dynamical evolution of dense stellar systems
containing up to $N\sim10^5-10^6$ stars in less than $\sim1\,$day
of computing time.

The evolution of globular clusters in the Galactic environment has
been studied using a variety of theoretical and numerical techniques.
The first comprehensive study of cluster lifetimes
was conducted by Chernoff \& Weinberg (1990, hereafter CW) using
F-P simulations.
They included the effects of a power-law mass spectrum, a tidal cut-off
radius imposed by the tidal field of the Galaxy, and mass loss due to stellar
evolution. Their results were surprising, and far reaching, since they
showed for the first time that the majority of clusters
with a wide range of initial conditions would be disrupted
in $\la 10^{10}\,$yr, and would not survive until core collapse.
CW carried out their calculations using a 1-D F-P method, in which
the stellar distribution function in phase space is assumed to depend
on the orbital energy only.
However, more recently, similar calculations undertaken using direct
\nbody\
simulations gave cluster lifetimes up to an order of magnitude longer
compared to those computed by CW (Fukushige \& Heggie 1995;
Portegies Zwart \etal 1998).
The discrepancy appears to be caused by an overestimated mass loss
rate in the 1-D F-P formulation (Takahashi \& Portegies Zwart 1998),
which does not properly account for the velocity anisotropy in the
cluster. To overcome this problem,
new 2-D versions of the F-P method (in which the distribution
function depends on both energy and angular momentum) have been employed
(Takahashi 1995, 1996, 1997; Drukier \etal 1999).

The 2-D F-P models provide cluster lifetimes in significantly better
agreement with direct \nbody\ integrations
(Takahashi \& Portegies Zwart 1998).
However, the 2-D F-P models still exhibit a slightly higher mass
loss rate compared to \nbody\ simulations. This may result from
the representation of the system in terms of a continuous distribution
function in the F-P formulation, which effectively models the behavior
of the cluster in the $N \to \infty$ limit. To test this possibility,
Takahashi \& Portegies Zwart (1998) introduced an additional free
parameter \nuesc\ in their F-P models, attempting to take into account
the finite ratio of the crossing time to the relaxation time
(see also Lee \& Ostriker 1987; Ross \etal 1997).
They used this free parameter to lower the overall mass loss rate
in their F-P models and obtained better agreement with \nbody\
simulations (performed with up to $N=32,768$).
Takahashi \& Portegies Zwart (2000, hereafter TPZ) show that,
after calibration, a single value of
\nuesc\ gives consistent agreement with \nbody\ simulations
for a broad range of initial conditions.

The first paper in this series  presented details about our new parallel
Monte Carlo code as well as the results of a series of initial
test calculations (Joshi, Rasio \& Portegies Zwart 2000, hereafter Paper~I).
We found excellent agreement between the results of our test calculations
and those of direct \nbody\ and 1-D Fokker-Planck simulations for
a variety of single-component clusters (i.e., containing equal-mass
stars).
However, we found that, for tidally truncated clusters, the mass
loss rate in our models was significantly lower, and the
core-collapse times significantly longer, than in corresponding
1-D F-P calculations.
We noted that, for a single case (a $W_0 = 3$ King model), our results
were in good agreement with those of 2-D F-P calculations by Takahashi (1999).

In this paper, we extend our Monte Carlo calculations to multi-component
clusters (described by a continuous, power-law stellar mass function),
and we study the evolution of globular clusters with a broad range of initial
conditions. Our calculations include an improved treatment of mass loss
through the tidal boundary, as well as mass loss due to stellar evolution.
Our new method treats the mass loss through the tidal boundary more
carefully in part by making the timestep smaller, especially in situations where
the tidal mass loss can lead to an instability resulting in rapid disruption
of the cluster. We also account for the shrinking of the tidal boundary
in each timestep by iteratively removing stars with apocenter distances greater
than the tidal boundary, and recomputing the tidal radius using the new (lower)
mass of the cluster.
We compare our new results with those of CW and TPZ.
We also go beyond these previous studies and
explore several other issues relating to the pre-collapse
evolution of globular clusters. We study in detail the importance
of the velocity anisotropy in determining
the stellar escape rate. We also compare the
orbital properties of escaping stars in disrupting and collapsing
clusters. Finally, we consider the effects of an eccentric orbit
in the Galaxy, allowing for the possibility that a cluster may
not fill its Roche lobe at all points in its orbit.

As in most previous studies, the calculations presented in this paper
are for clusters containing single stars only.
The dynamical effects of hard primordial binaries for the overall cluster
evolution are not significant during most of the
the {\em pre-collapse\/} phase, although a large primordial binary fraction
could accelerate the evolution to core collapse since binaries are on average
more massive than single stars. Energy generation
through binary -- single star and binary -- binary interactions becomes
significant only when the cluster approaches core collapse and
interaction rates in the core increase substantially
(Hut, McMillan \& Romani 1992; Gao \etal 1991; McMillan
\& Hut 1994). Formation of hard ``three-body'' binaries can also be
neglected until the cluster reaches a deep core-collapse phase.
During the pre-collapse evolution, hard binaries
behave approximately like single more massive stars,
while soft binaries (which have a larger
interaction cross section) may be disrupted.
Since we do not include the effects of energy generation by
primordial binaries in our
calculations, the (well-defined) core-collapse times presented here
may be re-interpreted as corresponding approximately to the
onset of the ``binary-burning'' phase, during which
a similar cluster containing binaries would be supported in
quasi-equilibrium by energy-generating interactions with hard
binaries in its core (Spitzer \& Mathieu 1980; Goodman \& Hut 1989;
McMillan, Hut \& Makino 1990, Gao \etal 1991).
Our calculations of disruption times (for clusters that
disrupt in the tidal field of the Galaxy before reaching core collapse)
are largely independent of the cluster binary content,
since the central densities and core interaction rates in these
clusters always remain very low.

Our paper is organized as follows.
In \S 2, we describe the treatment of tidal stripping and mass loss due
to stellar evolution in our Monte Carlo models, along with a discussion of
the initial conditions for our simulations.
In \S 3, we present the results of our simulations and comparisons
with F-P calculations. In \S 4, we summarize our results.

%%%%%%%%%%%%%%%%%%%%%%%%%%%%%%%%%%%%%%%%%%%%%%%%%%%%%%%%%%%%%%%%%%%%%%%%%%%

\section{Monte Carlo Method}

Our code, described in detail in Paper~I, is based
on the orbit-averaged Monte Carlo method first developed by Henon (1971a,b).
Although in Paper~I we only presented results of test calculations
performed  for single-component clusters,
the method is completely general, and the implementation of an arbitrary
mass spectrum is straightforward.
This section describes additional features of our code that
were not included in Paper~I: an improved treatment of mass loss
through the tidal boundary (\S2.1), and a simple implementation of
stellar evolution (\S2.2). The construction of initial multi-component
King models for
our study of cluster lifetimes is described in \S2.3. The highly
simplified treatments of tidal effects and stellar evolution
adopted here are for consistency with previous studies, since our
intent in this paper is still mainly to establish the accuracy of
our code by presenting detailed comparisons with the results of other
methods.
In future work, however, we intend to implement more sophisticated
and up-to-date treatments of these effects.

\subsection{Tidal Stripping of Stars}

In an isolated cluster, the mass loss rate (up to core collapse) is
relatively small, since escaping stars must acquire positive energies
mostly through rare, strong interactions in the dense cluster core (see
the discussion in Paper~I, \S 3.1). In contrast,
for a tidally truncated cluster, the mass loss is dominated by
diffusion across the tidal boundary (also referred to as ``tidal stripping'').
In our Monte Carlo simulations, a star is assumed to be tidally stripped from
the cluster (and lost instantaneously) if the {\em apocenter\/}
of its orbit in the cluster
is outside the tidal radius. This is in contrast to the {\em energy-based\/}
escape criterion that is used in 1-D F-P models, where a star is
considered lost if its energy is greater than the energy at the tidal
radius, regardless of its angular momentum.
As noted in Paper~I, the 2-D treatment is crucial in order to avoid
overestimating the escape rate, since stars with
high angular momentum, i.e., on more circular orbits, are less likely
to be tidally stripped from the cluster than those (with the same
energy) on more radial orbits.

A subtle, yet important aspect of the mass loss across the tidal boundary,
is the possibility of the tidal stripping process becoming {\em unstable\/}
if the tidal boundary moves inward too quickly. As the total mass of the
cluster decreases through the escape of stars, the tidal radius of the cluster
shrinks. This causes even more stars to escape, and the tidal boundary
shrinks further. If at any time during the evolution of the cluster the
density gradient at the tidal radius is too large, this can lead to an unstable
situation, in which the tidal radius continues to shrink on the dynamical
timescale, causing the cluster to disrupt. The development of
this instability characterizes the final evolution of all clusters with
a low initial central concentration that disrupt in the Galactic tidal
field before reaching core collapse.

We test for this instability at each timestep in our simulations, by
iteratively removing escaping stars and recomputing the tidal radius
with the appropriately lowered cluster mass. For stable models, this
iteration converges quickly, giving a finite escape rate.
Even before the development of the instability, this iterative
procedure must be used for an accurate determination of the mass loss
rate. When the mass loss rate due to tidal stripping is high,
we also impose a timestep small enough that no more than 1\% of
the total mass is lost in a single timestep. This is to ensure
that the potential is updated frequently enough to take the mass
loss into account.
This improved treatment of tidal stripping was not used in our
calculations for Paper I. However, all the results presented
in Paper I were for clusters with equal-mass stars, with no stellar
evolution. Under those conditions, all models reach core collapse, with
no disruptions. The issue of unstable mass loss is not significant
in those cases, and hence the results of Paper I are unaffected.

\subsection{Stellar Evolution}

Our simplified treatment follows  those adopted by CW and TPZ.
We assume that a star evolves instantaneously
to become a compact remnant at the end of its main-sequence lifetime.
Indeed, since the evolution of our cluster models takes place
on the relaxation timescale (i.e., the timestep is a fraction
of the relaxation time $t_r\ga 10^9\,$yr), while the dominant mass loss phase
during late stages of stellar evolution takes place on a much shorter
timescale
($\sim 10^6\,$yr), the mass loss can be considered instantaneous.
We neglect mass losses in stellar winds for main-sequence stars.
We assume that the main-sequence lifetime and remnant mass is a function
of the initial stellar mass only. Table~1 shows the
main-sequence lifetimes of stars with initial masses up to $15\,\msun$, and
the corresponding remnant masses. In order to facilitate comparison with
F-P calculations (CW, and TPZ), we use the same lifetimes and remnant masses
as CW. For stars of mass $m < 4\,\msun$, the remnants are white dwarfs
of mass $0.58\,\msun + 0.22\,(m-\msun)$, while for $m > 8\,\msun$,
the remnants are neutron stars of mass $1.4\,\msun$.
Stars with intermediate masses
are completely destroyed (Iben \& Renzini 1983).
The lowest initial mass considered by CW was $\simeq 0.83\,\msun$.
For lower mass stars, in order to maintain consistency with TPZ, we
extrapolate the lifetimes assuming a simple $m^{-3.5}$ scaling
(Drukier 1995).
We interpolate the values given in Table~1 using a cubic spline to
obtain lifetimes for stars with intermediate masses, up to $15\,\msun$.
In our initial models (see \S2.3), we assign masses to stars according
to a continuous
power-law distribution. This provides a natural spread in their lifetimes,
and avoids having large numbers of stars undergoing identical stellar
evolution. In contrast, in F-P calculations,
the mass function is approximated by $20$ discrete logarithmically spaced
mass bins over the entire range of masses. The mass in each bin is then
reduced linearly in time from its initial mass to its final (remnant)
mass, over a time interval equal to the maximum difference in main-sequence
lifetimes spanned by the stars in that mass bin (see TPZ for further
details). This has the effect of averaging the effective mass loss
rate over the masses in each bin.

We assume that all stars in the cluster were formed in the same star
formation epoch, and hence all stars
have the same age throughout the simulation.
During each timestep, all the stars that have evolved beyond their
main-sequence lifetimes are labelled as remnants, and their masses are
changed accordingly. In the initial stages of evolution ($t\la 10^8\,$yr),
when the mass loss rate due to stellar evolution is highest, care is taken to
make the timestep small enough so that no more than 1\% of the total
mass is lost in a single timestep. This is to ensure that the system
remains very close to virial equilibrium through this phase.

\subsection{Initial Models}

The initial condition for each simulation is a
King model with a power-law mass spectrum. In order to facilitate comparison
with the F-P calculations of CW and TPZ, we select the same set of
initial King models for our simulations, with values of the
dimensionless central potential $W_0 =\,$1, 3, and~7.
Most of our calculations were performed with $N=10^5$ stars, with
a few calculations repeated with $N=3\times10^5$ stars and showing no
significant differences in the evolution.
We construct the initial model by first generating a single-component
King model with the selected $W_0$. We then assign
masses to the stars according to a power-law mass function
\begin{equation}
f(m) \propto m^{-\alpha},
\end{equation}
with $m$ between $0.4\,\msun$ and $15\,\msun$.
We consider three different values for
the power-law index $\alpha =\,$1.5, 2.5, and~3.5, assuming no
initial mass segregation.
Although this method of generating a multi-component initial
King model is convenient and widely used to create initial conditions
for numerical work (including \nbody\ , F-P,
and Monte Carlo simulations), the resulting initial model is not
in strict virial equilibrium since the masses are assigned independently
of the positions and velocities of stars. However, we find that the
initial clusters
relax to virial equilibrium within just a few timesteps in our
simulations. Virial equilibrium is then maintained to high accuracy
during the entire calculation, with the virial ratio $2T/|W|=1$ to
within $< 1$\%.

In addition to selecting the dimensionless model parameters $W_0$, $N$,
and $\alpha$ (which specify the initial dynamical state of the system),
we must also relate the dynamical timescale with the stellar evolution
timescale for the system. The basic unit of time in our models is
scaled to the relaxation time.
Since the stellar evolution timescale is not directly related to the
dynamical timescale, the lifetimes
of stars (in years) cannot be computed directly from our code units.
Hence, in order to compute the mass loss due to stellar evolution,
we must additionally relate the two timescales by converting the evolution
time to physical units. To maintain consistency with F-P calculations,
we use the same prescription as CW. We assume a value for the initial
relaxation time of the system, which is defined as follows:
\begin{equation}
        t_{\rm r} = 2.57\,F\;  [{\rm Myr}],
\end{equation}
where
\begin{equation}
        F \equiv \frac{M_0}{\msun}\frac{R_{\rm g}}{\rm kpc}
                 \frac{\rm 220~km\,s^{-1}}{v_{\rm g}}
                 \frac{1}{\ln N} .
\end{equation}
Here $M_0$ is the total initial mass of the cluster, $R_{\rm g}$ is its
distance to the Galactic center (assuming a circular orbit),
$v_{\rm g}$ is the circular speed of the
cluster, and $N$ is the total number of stars.  (This expression for
the relaxation time is derived from CW's eqs.~[1], [2], and [6]
with $m=\msun$, $r=r_{\rm t}$, and $c_1=1$.)
Following CW, a group of models with the same value of $F$
(constant relaxation time) at the beginning of the simulation
is referred to as a ``Family.''  Our survey covers CW's
Families~1, 2, 3 and~4. For each value of $W_0$ and $\alpha$,
we consider four different models, one from each Family.

To convert from our code units, or ``virial units'' (see Paper~I,
\S 2.8 for details) to physical units, we proceed as follows.
For a given Family (i.e, a specified value of $F$), cluster mass $M_0$,
and $N$, we compute the distance to the Galactic center $R_g$ using
equation~(3). The circular velocity of $220\,\kms$ for the cluster
(combined with $R_g$) then provides an inferred value for the mass of
the Galaxy $M_g$ contained within the cluster orbit. Using $M_0$, $M_g$,
and $R_g$, we compute the tidal radius for the cluster, as
$r_t = R_g\, (M_0 / 3 M_g)^{1/3}$, in physical units (pc).
The ratio of the tidal radius to the virial radius (i.e., $r_t$
in code units) for a King model depends only on $W_0$, and hence is
known for the initial model. This gives the virial radius in pc.
The unit of mass is simply the total initial cluster mass
$M_0$. Having expressed the units of distance and mass in physical units,
the unit of evolution time (which is proportional to the relaxation
time) can easily be converted to physical units (yr) using
equation~(31) from Paper I.

Table~2 shows the value of $F$ for the four selected Families.
For reference, we also give the relaxation time at the half-mass
radius $t_{\rm rh}$ for the models with $W_0=3$ and $\alpha=2.5$
(mean stellar mass $\bar{m} \simeq 1\,\msun$), which we compute
using the standard expression (see, e.g., Spitzer 1987),
\begin{equation}
        t_{\rm rh} = 0.138\frac{N^{1/2}r_{\rm h}^{3/2}}
                               {\bar{m}^{1/2}G^{1/2} \ln N},
\end{equation}
where $r_{\rm h}$ is the half-mass radius of the cluster.

%%%%%%%%%%%%%%%%%%%%%%%%%%%%%%%%%%%%%%%%%%%%%%%%%%%%%%%%%%%%%%%%%%%%%%%%%%%

\section{Results}

In Paper~I we presented our first results for the evolution of
single-component clusters up to core collapse. We computed core-collapse
times for the entire sequence of King models ($W_0 = 1-12$), including
the effects of a tidal boundary. Here we extend our study to clusters
with a power-law mass spectrum, and mass loss due to stellar evolution.

\subsection{Qualitative Effects of Tidal Mass Loss and Stellar Evolution}

We begin by briefly reviewing the evolution of single-component, tidally
truncated systems. In Figure~1, we show the core-collapse times for
King models with $W_0 = 1-12$ (Paper~I).
The core-collapse times for tidally truncated
models are compared with equivalent isolated models. Although the isolated
models also begin as King models with a finite tidal radius, the tidal boundary
is not enforced during their evolution, allowing the cluster to expand freely.
The most notable result is that the maximum
core-collapse time for the tidally truncated clusters occurs at $W_0
\simeq 5$, compared to $W_0 = 1$ for isolated clusters.
This is because the low $W_0$ King models have a less centrally concentrated
density profile, and hence a higher density at the tidal radius compared
to the high $W_0$ models. This leads to higher mass loss through the
tidal boundary, which reduces the mass of the cluster and shortens the
core-collapse time.
This effect is further complicated by the introduction of a non-trivial
mass spectrum, and mass loss due to stellar evolution in the cluster.

In Figure~2, we show a comparison of the mass loss rate due to the
tidal boundary, a power-law mass spectrum, and stellar evolution.
We consider the evolution of a $W_0 = 3$ King model, in four different
environments.
All models considered in this comparison belong to Family~1 (cf.\ \S 2.3).
We first compare an isolated, single-component model
(without an enforced tidal boundary), and a tidally truncated model
(as in Fig.~1). Clearly, the presence of the tidal boundary
is responsible for almost all the mass loss from the cluster, and
it slightly reduces the core-collapse time.
Introducing a power-law mass spectrum further reduces the core-collapse
time, since mass segregation increases the core density, and accelerates
the development of the gravothermal instability.
The shorter core-collapse time reduces the total mass loss through
the tidal boundary by leaving less time for evaporation.
This results in a higher final mass compared to
the single-component system, even though the mass loss rate is
higher. Finally, allowing mass loss through stellar evolution causes
even faster overall mass loss, which eventually disrupts the system.
The introduction of
a Salpeter-like power-law initial mass function ($\alpha = 2.5$) is
sufficient to cause this cluster to disrupt before core collapse.

The presence of a tidal boundary causes stars on radial orbits in
the outer regions of the cluster to be preferentially removed. This
produces a significant anisotropy in the outer regions
as the cluster evolves. As noted in Paper~I, a proper
treatment of this anisotropy is essential in computing the mass
loss rate.
A star in an orbit with low angular momentum has a larger apocenter
distance compared to a star (with the same energy) in a high angular
momentum orbit.
Hence stars in low angular momentum (i.e., radial) orbits are
preferentially lost through the tidal boundary, causing an
anisotropy to develop in the cluster.
In 1-D F-P models, this is not taken into account, and therefore
1-D F-P models predict a much larger mass loss compared to 2-D models.
In Figure~3, we show the anisotropy parameter
$\beta = 1 - \sigma_t^2/\sigma_r^2$, for a $W_0 = 3$ King model
($\alpha = 2.5$, Family~1), at two different times during its evolution.
Here, $\sigma_t$ and $\sigma_r$ are the 1-D tangential and
radial velocity dispersions, respectively. The
initial King model is isotropic. At later times, the anisotropy
in the outer region grows steadily as the tidal radius
moves inwards.

Another important consequence of stellar evolution and mass segregation
is the gradual flattening of the stellar mass function as the
cluster evolves. In Figure~4, we show the main-sequence
mass spectrum in the
core and at the half-mass radius of a $W_0 = 7$ King model
($\alpha = 2.5$, Family~2), at two different times during its
evolution.
Since the heavier stars concentrate in the core, and have lower mean
velocities, the mass loss across the tidal boundary occurs preferentially
for the lighter stars. This leads to a gradual flattening of the
overall mass function of the cluster.
However, this picture is somewhat complicated by stellar evolution,
which continuously depletes high-mass stars from the cluster.
The remaining heavier stars gradually accumulate in the inner
regions as the cluster evolves. Therefore the flattening of the
mass function becomes particularly evident in the cluster core.

\subsection{Cluster Lifetimes: Comparison with Fokker-Planck results}

We now present our survey of cluster lifetimes, and we compare our
results with equivalent 1-D and 2-D F-P results.
For each combination of $W_0$ and $\alpha$, we perform four different
simulations (Families~$1-4$), corresponding to different initial
relaxation times (cf. Table~[2]). We follow the evolution until core
collapse, or disruption, whichever occurs first. We also stop the
computation if the total bound mass decreases below 2\% of the initial
mass, and consider the cluster to be disrupted in such cases.
We compare our results with those of two different F-P studies:
the 1-D F-P calculations of Chernoff \& Weinberg (1990, CW), and
the more recent 2-D calculations of Takahashi \& Portegies Zwart (2000, TPZ).

\subsubsection{Comparison with 1-D \FP models}

Table~3 compares the our Monte Carlo (MC) models with the 1-D F-P
calculations conducted by CW.
Following the same notation as CW, the final core collapse of a
cluster is denoted by `C', and disruption is denoted by `D'.
The final mass of the cluster (in units of the initial mass)
and the lifetime in units of $10^9$ yr (time to disruption or core
collapse) are also given.
The evolution of clusters that reach core collapse is not followed
beyond the core-collapse phase.
The core-collapse time is taken as the time when the
the innermost lagrange radius (radius containing 0.3\% of the
total mass of the cluster) becomes smaller than 0.001 (in
virial units).
For disrupting clusters, CW provide a value for the final mass,
which corresponds to the point at which the tidal mass
loss becomes unstable and the cluster disrupts on the dynamical
timescale. However, we find that the point at which the instability
develops depends sensitively on the method used for computing the tidal
mass loss and requires the potential to be updated on a very short
timescale. In this regime,
since the system evolves (and disrupts) on the dynamical timescale,
the orbit-averaged approximation used to solve the Fokker-Planck
equation also breaks down. This is true for both Monte Carlo
and F-P simulations. The only way to determine the
point of instability reliably is to follow the evolution on the
dynamical timescale using direct \nbody\ integrations.
Hence, for disrupting models, we quote the final mass as zero,
and only provide the disruption time (which can be determined
very accurately).

We find that {\em all\/} our Monte Carlo models disrupt later
than those of CW.  However, for models that undergo core collapse,
the core-collapse times are shorter in some cases compared to CW,
because the lower mass loss rate in our Monte Carlo models
causes core collapse to take place earlier.
The discrepancy in the disruption times sometimes
exceeds an order of magnitude (e.g., $W_0=1$, $\alpha=2.5$).
On the other hand, the discrepancy in the lifetimes of
the clusters with $\alpha=1.5$, $W_0=1$ \& 3 is
is only about a factor of two.  These models disrupt very
quickly and a proper treatment of anisotropy does not extend their
lifetimes very much, since the combination of a flat initial mass
function and a shallow initial potential leads to rapid disruption.

Out of 36 models, we find that half (18) of our Monte Carlo models
reach core collapse before disruption, compared to fewer than
30\% (10) of models in the CW survey.  The longer lifetimes of our
models allow more of the clusters to reach core collapse in our
simulations.
All the clusters that experience core collapse according to CW also
experience core collapse in our calculations. Since the main difference
between our models and those of CW comes from the different mass loss
rates, we predictably find that our results match more closely those of
CW in all cases where the overall mass loss up to core collapse is
relatively small. For example, the more concentrated clusters ($W_0 = 7$) with
steep mass functions ($\alpha=$2.5 and 3.5) show very similar behavior,
with the discrepancy in final mass and core-collapse time being less
than a factor of two.
However, we cannot expect complete agreement even in these
cases, since the effects of anisotropy cannot be completely ignored.

The overall disagreement between our Monte Carlo models and 1-D F-P
models is very significant. This was also evident in some of the
results presented in Paper~I, where
we compared core-collapse times for tidally truncated single-component King
models, with 1-D F-P calculations by Quinlan (1996). This discrepancy
has also been noted by Takahashi \& Portegies Zwart (1998), and Portegies
Zwart \etal (1998). The improved 2-D F-P code developed by Takahashi
(1995, 1996, 1997) is now able to properly account for the anisotropy,
allowing for a more meaningful comparison with other 2-D calculations,
including our own.

\subsubsection{Comparison with 2-D \FP models}

Comparisons of the mass loss evolution is shown in
Figures~5, 6, and 7, where the solid lines show our Monte Carlo models,
and the dashed lines show the 2-D F-P models from TPZ.

In Figure~5, we show the evolution of $W_0 = 1$ King models.
The very low initial central density of these models makes them very
sensitive to the tidal boundary, leading to very rapid mass loss.
As a result, almost all the $W_0 = 1$ models  disrupt
without ever reaching core collapse.
In addition, these models demonstrate the largest variation in
lifetimes depending on their initial mass spectrum.
For a relatively flat mass function ($\alpha = 1.5$),
the disruption time is $\sim 2\times 10^7\,$yr. The large fraction
of massive stars in these models, combined with the shallow initial
central potential, leads to very rapid mass loss and complete
disruption. For a more
realistic, Salpeter-like initial mass function ($\alpha = 2.5$),
the $W_0 = 1$ models have a longer lifetime, but still disrupt in
$\la 10^9\,$yr. The $\alpha = 3.5$ models have very few massive
stars, and hence behave almost like models without stellar evolution.
We see that it is only with such a steep mass function, that
the $W_0 = 1$ models can survive until the present epoch
($\ga 10^{10}\,$yr). We also find that the Family~1 and~2 models
can reach core collapse despite having lost most of their
mass, while Family~3 and~4 models are disrupted.

We see very good agreement throughout the evolution between our
Monte Carlo models and the 2-D F-P models.
In all cases, the qualitative behaviors indicated by the two methods
are identical, even though the Monte Carlo models consistently
have somewhat longer lifetimes than the F-P models. The average
discrepancy in the disruption times for all models is
approximately a factor of two. The discrepancy
in disruption times is due to a slightly lower mass loss rate in
our models, which allows the clusters to live longer. Since the F-P
calculations correspond to the $N \to \infty$ limit, they tend to
overestimate the overall mass loss rate
(we discuss this issue in more detail in the next section).
This tendency has been pointed out by Takahashi \& Portegies Zwart
(1998), who compared the results of 2-D F-P simulations with
those of direct $N$-body simulations with up to $N=32,768$.
They have attempted to account for the finiteness of the system in
their F-P models by introducing an additional parameter in their
calculations to modify the mass loss rate.
The comparison shown in Figures~5, 6 and~7
is for the unmodified $N \to \infty$ F-P models.

We find complete agreement with TPZ
in distinguishing models that reach core collapse from those that
disrupt. The only case in which there is some ambiguity is
the $W_0 =1, \alpha = 3.5$, Family~2 model, which clearly collapses in our
calculations, while TPZ indicate nearly complete disruption. This is
obviously a
borderline case, in which the cluster reaches core collapse just prior
to disruption in our calculation. Since the cluster has lost almost
all its mass at core collapse, the distinction between core collapse
and disruption is largely irrelevant. It is important to note, however,
that we find the boundary between collapsing and disrupting models
at almost exactly the same location in parameter space
($W_0$, $\alpha$, and relaxation time) as TPZ. This agreement
is as significant, if not more, than the comparison of final masses
and disruption times.

In Figure~6, we show the comparison of $W_0 = 3$ King models.
Again, the overall agreement is very good, except for the slightly later
disruption times for the Monte Carlo models.
The most notable difference from the $W_0 = 1$
models, is that the $W_0 = 3$ models clearly reach core collapse
prior to disruption for $\alpha = 3.5$. The core collapse times
for the $\alpha = 3.5$ models are very long ($3\times 10^{10} - 3\times 10^{11}\,$yr),
with only $\sim 20\%$ of the initial mass remaining bound at
core collapse. Here also we find perfect agreement between the qualitative
behaviors of the F-P and Monte Carlo models.

In Figure~7, we show the evolution of the $W_0 = 7$ King models.
In the presence of a tidal boundary, the $W_0 \simeq 5$ King models
have the distinction of having the longest core-collapse times
(see Fig.~1).
This is because they begin with a sufficiently high initial core
density, and do not expand very much before core collapse. Hence,
the mass loss through the tidal boundary is minimal. King models
with a lower $W_0$ lose more mass through the tidal boundary,
and evolve more quickly toward core collapse or disruption,
while models with higher $W_0$ have very high initial core densities,
leading to short core-collapse times. All our $W_0 = 7$ models reach
core collapse. Even the models with a very flat mass function
($\alpha = 1.5$) achieve core collapse, although the final bound
mass in that case is very small. We again see very good overall agreement
between the Monte Carlo and F-P models, except for the slightly
higher mass loss rate predicted by the F-P calculations.
In the next section, we discuss the possible reasons for this small
discrepancy in the mass loss rate between the Monte Carlo and
F-P models.

\subsubsection{Comparison with finite \FP models}

We first highlight some of the general issues relating to mass loss
in the systems we have considered.
In Figure~8, we show the relative rates of mass loss due to
stellar evolution and tidal stripping, for $W_0 = $1, 3, and 7
King models, with different mass spectra
($\alpha = $ 1.5, 2.5, and 3.5).
We see that stellar evolution is most significant
in the early phases, while tidal mass loss dominates the evolution
in the later phases. The relative importance of stellar evolution
depends on the fraction of massive stars in the cluster, which
dominate the mass loss early in the evolution. Hence,
the $\alpha = 1.5$ models suffer the greatest mass loss due
to stellar evolution, accounting for up to 50\% of the total mass loss
in some cases (e.g., $W_0 = 7, \alpha = 1.5$). All models
shown belong to Family~2.
It is important to note the large variation in the timescales, and
in the relative importance of stellar evolution versus tidal mass loss
across all models.

Through comparisons with $N$-body simulations,
Takahashi \& Portegies Zwart (1998) have argued that assuming
$N \to \infty$ leads to an overestimate of the mass loss
rate due to tidal stripping of stars.
To compensate for this, they introduce a free parameter
\nuesc\ in their calculations, to account for the finite
time (of the order of the crossing time) it takes for an escaping
star to leave the cluster. They calibrate this parameter through
comparisons with \nbody\ simulations, for $N=1,024 - 32,768$).
Since for low $N$, the \nbody\ models are too noisy, and the
F-P models are insensitive to \nuesc\ for large $N$, TPZ
find that the calibration is most suitably done using
$N \sim 16,000$ (for further details, see the discussion by TPZ).
They show
that a single value of this parameter gives good agreement with
$N$-body simulations for a wide range of initial conditions.
Using this prescription, TPZ provide results of their calculations
for finite clusters with $N = 3\times 10^5$ in addition to their
$N \to \infty$ results. They find that their finite models,
as expected, have lower mass loss rates, and consequently longer
lifetimes compared to their infinite models.

In Table~4, we compare the results of our Monte Carlo calculations
with $N = 3\times 10^5$ stars with the finite and infinite F-P
models of TPZ. We consider
four cases: $W_0 = $ 1 and 3, Families~1 and 4, $\alpha = 2.5$.
All finite TPZ models have longer lifetimes than their
infinite models. However, there is practically no difference
between their finite and infinite models for core-collapsing
clusters. Hence we focus our attention only on the disrupting
models. We see that in all four cases, the longer lifetimes
of the finite models are in better agreement with our
Monte Carlo results, although the agreement is still not
perfect. The largest difference between the finite and
infinite F-P models is for the $W_0 = 1$ models, in which
case the Monte Carlo results lie between the finite
and infinite F-P results. For $W_0 = 3$ models, the
Monte Carlo disruption times are still slightly longer than those of
the finite F-P models, although the agreement is better.

Both Monte Carlo and F-P methods are based on the
orbit-averaged Fokker-Planck approximation, which treats all
interactions in the weak scattering limit, i.e., it does not take
into account the effect of strong encounters. Both methods compute
the \emph{cumulative} effect of distant encounters in one timestep
(which is a fraction of the relaxation time). In this
approximation, events on the crossing timescale (such as the escape
of stars) are treated as being instantaneous. Since the relaxation
time is proportional to $N/\ln N$ times the crossing time, this is
equivalent to assuming $N \to \infty$ in the F-P models. However,
in our Monte Carlo models, there is \emph{always} a finite $N$,
since we maintain a discrete representation of the cluster at
all times and follow the same phase space parameters as in an
\nbody\ simulation.
Thus, although both methods make the same assumption
about the relation between the crossing time and relaxation time,
for all other aspects of the evolution, the Monte Carlo models
remain finite. This automatically allows most aspects of cluster
evolution, including the escape of stars, stellar evolution, and
computation of the potential, to be handled on a discrete, star-by-star
basis. On the other hand, the F-P models use a few coarsely binned
individual mass components represented by continuous distribution
functions (consistent with $N \to \infty$) to model all processes.
In this sense, the Monte Carlo models can be regarded
as being intermediate between direct \nbody\ simulations and
F-P models.

The importance of using the correct value of $N$ in dynamical
calculations for realistic cluster models has also been
demonstrated through \nbody\ simulations, which show that the
evolution of finite clusters scales with $N$ in a rather complex way
(see Portegies Zwart \etal 1998 and the ``Collaborative Experiment''
by Heggie \etal 1999). Hence, despite correcting
for the crossing time, it is not surprising that the finite
F-P models are still slightly different from the Monte Carlo models.
It is also possible that the calibration of the escape parameter
obtained by TPZ may not be applicable to large $N$ clusters,
since it was based on comparisons with smaller $N$-body simulations.
It is reassuring to note, however, that the Monte Carlo
models, without introducing any new free parameters,
have consistently lower mass loss rates compared to the
infinite F-P models, and show better agreement with the
finite F-P models.

\subsection{Velocity and Pericenter Distribution of Escaping Stars}

A major advantage of the Monte Carlo method is that it allows the evolution
of specific subsets of stars, or even individual stars, to be followed
in detail. We use this capability to examine, for the first time
in a cluster simulation with realistic $N$, the
properties of stars that escape from the cluster through tidal stripping.
We also examine the differences between the properties of
escaping stars in clusters that reach core collapse, and those that
disrupt. In Figure~9, we show the distribution of escaping stars for
two different models ($W_0 = 3$ and 7, Family~1, $\alpha = 2.5$).
In each case, we show a 2-D distribution of the pericenter distance
and the velocity at infinity for all the escaping stars.
The velocity at infinity is computed as $v_{\infty} = \sqrt{2(E-\phi_t)}$,
where $E$ is the energy per unit mass of the star, and $\phi_t$ is the
potential at the tidal radius.
We see that the distribution
of pericenter distances is very broad, indicating that escape takes place
from within the entire cluster, and not just near the tidal boundary.
We see that the distribution of pericenters is slightly more centrally
peaked in the $W_0 = 7$ model than in the $W_0 = 3$ case. Note that
the sizes of the cores are very different for the two clusters.
The $W_0 = 7$ cluster initially has a core radius of 0.2 (in virial
units), which gets smaller as the cluster evolves, while the $W_0 = 3$
cluster has an initial core radius of 0.5, which does not
change significantly as the cluster evolves and disrupts.
The main difference between the clusters, however, is in the
velocity distribution of escaping stars.
In the disrupting cluster ($W_0 = 3$),
the escaping stars have a wide range of escape energies at all pericenter
distances, whereas in the
collapsing cluster ($W_0 = 7$), a large fraction of the stars escape with
close to the minimum energy. Only the escapers from within the
central region have a significant range of escape energies.

The very narrow distribution of escape energies for the collapsing
cluster suggests that the mechanism for escape in collapsing and
disrupting clusters may be qualitatively different.
It also suggests that the single escape
parameter used by TPZ to correct for the tidal mass loss rate
in their finite F-P calculations may be insufficient in correcting
for both types of escaping stars. This might also account for the
fact that TPZ find almost no change in the mass loss rate after
introducing their \nuesc\ parameter in core-collapsing models,
while disrupting models show a significant difference.

\subsection{Effects of Non-circular Orbits on Cluster Lifetimes}

In all the calculations presented above (as in most previous
numerical studies of globular cluster evolution), we assumed that the
cluster remained in a circular orbit at a fixed distance from the center
of the Galaxy.
We also assumed that the cluster was born filling its Roche lobe in
the tidal field of the Galaxy. Both of these assumptions are
almost certainly unrealistic for the majority of clusters. However, one
could argue that even for a cluster on an eccentric orbit, one
might still be able to model the evolution using an appropriately
\emph{averaged} value of the tidal radius over the orbit of the
cluster. Here we briefly explore the effect of an eccentric orbit,
by comparing the evolution of one of our Monte Carlo models
($W_0 = 3$, $\alpha = 2.5$, Family~2) on a Roche-lobe filling
circular orbit, and on an eccentric orbit. We assume that the
\emph{pericenter} distance of the eccentric orbit is equal to
the radius of the circular orbit. This is to ensure that
the cluster fills its Roche lobe at the same location,
and the same value of $R_g$ is used to compute $F$
in the models being compared (see eq.~[3]).
If we alternatively selected the orbit such that the cluster
fills its Roche lobe at apocenter, instead of pericenter, the
outcome would be obvious: the mass loss at pericenter would be
considerably higher, leading to much more rapid disruption of
the cluster compared to the circular orbit.

In Figure~10, we show the
evolution of the selected model for three different orbits.
The leftmost line shows the evolution for the circular orbit.
The rightmost line shows the evolution for an eccentric
\emph{Keplerian} orbit with a typical eccentricity of 0.6
(see, e.g., Odenkirchen \etal 1997). The Keplerian
orbit assumes that the inferred mass of the Galaxy interior to
the circular orbit is held fixed for the eccentric orbit as well.
The intermediate line shows the evolution for an orbit in
a more realistic potential for the Galaxy, which is still
spherically symmetric, but with a constant circular velocity
of $220 \kms$ in the region of the cluster orbit
(Binney \& Tremaine 1987). The orbit is chosen so that
it has the same pericenter and apocenter distance as the
Keplerian orbit. However, since the orbital velocity is higher,
it has a shorter period compared to the Keplerian orbit.
In each of the two eccentric orbits, we see that the cluster
lifetime is extended slightly (by a factor of $\sim 2$).
Most of the mass loss takes place
during the short time that the cluster spends near its pericenter,
where it fills its Roche lobe. The Keplerian orbit gives the
longest lifetime, since the cluster spends most of its time near
its apocenter, where it does not fill its Roche lobe.

This comparison suggests that the lifetime of a cluster can
vary by at most a factor of a few, depending on the shape of
its orbit. However, such corrections should
be taken into account in building accurate numerical models
of real clusters. In addition, other effects that we have neglected
here, such as tidal shocking during Galactic disk crossings, may
affect cluster lifetimes more significantly (see \S4).

\section{Summary}

We have calculated lifetimes of globular clusters in the Galactic
environment using 2-D Monte Carlo simulations with
$N = 10^5 - 3\times 10^5$ King models, including the effects
of a mass spectrum, mass loss in the Galactic tidal field,
and stellar evolution. We have studied the evolution of King models with
$W_0=1$, 3, and~7, and with power-law mass functions $m^{-\alpha}$, with
$\alpha=\,$1.5, 2.5, and~3.5, up to core collapse, or disruption,
whichever occurs first.
In our broad survey of cluster lifetimes,
we find very good overall agreement between our Monte Carlo models
and the 2-D F-P models of TPZ
for all 36 models studied. This is very reassuring,
since it is impossible to verify such results using direct \nbody\
integrations for a realistic number of stars.
The Monte Carlo method has been shown to be a robust alternative for
studying the evolution of multi-component clusters. It is particularly
well suited to studying finite, but large-$N\,$ systems, including
many different processes, such as tidal stripping and stellar evolution,
which operate on different timescales.
We find that our Monte Carlo models are in better agreement with the
finite-$N\,$ F-P models of TPZ, compared to their standard F-P
($N \to \infty$) models, although our models still appear to have
a slightly lower overall mass loss rate.

Even though our simulations are becoming more sophisticated and realistic
with the inclusion of many new important processes, there still
remain substantial difficulties in relating our results directly
to observed clusters. We ignore several potentially important effects in
these calculations, including the tidal shock heating of the cluster
following passages through the Galactic disk, and the presence of
primordial binaries, which can support the core against collapse.
In recent studies using 1-D F-P calculations, it has been shown that
shock heating and shock-induced relaxation of clusters caused by
repeated close passages near the bulge and through the disk of the
Galaxy can sometimes be as important as two-body relaxation for their
overall dynamical evolution (Gnedin, Lee, \& Ostriker 1999).
In addition, the initial mass function of clusters is poorly
constrained observationally, and our simple power laws may not be
realistic. In our study, we assume that clusters begin their
lives filling their Roche lobes. But, as we have shown, a cluster on an
eccentric orbit may spend most of its time further away in the
Galaxy, where it might not fill its Roche lobe. This can lead to
somewhat longer lifetimes.

The broad survey of cluster lifetimes presented here, and the similar
effort by TPZ, lay the foundations for more detailed calculations,
which may one day allow us to conduct reliable population synthesis
studies to understand in detail the history, and predict the future
evolution, of the Galactic globular cluster system.

\acknowledgements

We are very grateful to Simon Portegies Zwart for insightful
comments and helpful discussions. We are also grateful to Koji Takahashi
for kindly providing valuable data and answering numerous questions.
This work was supported by NSF Grant AST-9618116 and NASA ATP
Grant NAG5-8460. C.P.N.\ acknowledges partial support from the
UROP program at MIT. F.A.R.\ was supported in part by an Alfred P.\
Sloan Research Fellowship. This
work was also supported by the National Computational Science Alliance
under Grant AST980014N
and utilized the SGI/Cray Origin2000 supercomputer at Boston University.

\clearpage

% -----------------------------  Tables  --------------------- %

\begin{deluxetable}{ccc}
\tablecaption{Main-Sequence Lifetimes and Remnant Masses $^{\rm a}$}
\tablewidth{0pt}
\tablehead{
\colhead{$m_{initial} [\msun]$} &
\colhead{$\log(\tau_{\rm MS} [{\rm yr}])$} &
\colhead{$m_{final} [\msun]$} \\
}
\startdata
0.40 & 11.3 & 0.40  \\
0.60 & 10.7 & 0.49  \\
0.80 & 10.2 & 0.54  \\
1.00 & 9.89 & 0.58  \\
2.00 & 8.80 & 0.80  \\
4.00 & 7.95 & 1.24  \\
8.00 & 7.34 & 0.00  \\
15.00 & 6.93 & 1.40 \\
\enddata
\tablenotetext{a}{For consistency, we use the same main-sequence lifetimes and
remnant masses as CW, from Iben \& Renzini (1983) and Miller \& Scalo (1979).}
\end{deluxetable}

\begin{deluxetable}{lccc}
\tablecaption{Family Properties $^{\rm a}$}
\tablewidth{0pt}
\tablehead{
\colhead{Family} & \colhead{$F$}
& \colhead{$t_{\rm rh}$ [Gyr]} & \colhead{$R_g$ [kpc]}
}
\startdata
1 & $5.00 \times 10^4$ & 2.4 & 5.8  \\
2 & $1.32 \times 10^5$ & 6.4 & 15 \\
3 & $2.25 \times 10^5$ & 11 & 26  \\
4 & $5.93 \times 10^5$ & 29 & 68  \\
\enddata
\tablenotetext{a}{Sample parameters for Families~1--4,
for a $W_0 = 3$ King model, with $\bar m = 1 \msun$, and $N = 10^5$.
Distance to the Galactic center $R_g$ is computed assuming
that the cluster is in a circular orbit, filling its Roche
lobe at all times.}
\end{deluxetable}

\begin{deluxetable}{cccccccccc}
\tablecaption{Comparison of Monte Carlo results with 1-D \FP calculations $^{\rm a}$}
\tablewidth{0pt}
\scriptsize
%\footnotesize
\tablehead{
\colhead{} & \colhead{} & \multicolumn{8}{c}{Family} \\
\cline{3-10}
\colhead{$W_0$} & \colhead{$\alpha$} & \multicolumn{2}{c}{1} &
\multicolumn{2}{c}{2} &
\multicolumn{2}{c}{3} & \multicolumn{2}{c}{4}  \\
\cline{3-10}
\colhead{} & \colhead{} & \colhead{CW} & \colhead{MC} & \colhead{CW} &
\colhead{MC} & \colhead{CW} & \colhead{MC} & \colhead{CW} & \colhead{MC}
}
\startdata
%& & \phantom{AAAAAA} & \phantom{AAAAAA} & \phantom{AAAAAA}
%& \phantom{AAAAAA}
%& \phantom{AAAAAA} & \phantom{AAAAAA} & \phantom{AAAAAA}
%& \phantom{AAAAAA} \\
%
%                          CW      MC      CW     MC      CW       MC     CW       MC
%--------------------------------------------------------------------------------------
1 ............  & 1.5   & D     & D     & D     & D     & D     & D     & D     & D     \\
                &       &0.0092 & 0.019  &0.0094 & 0.019  &0.0093 & 0.02  &0.0092 & 0.02  \\
                &       & 0     & 0     & 0     & 0     & 0     & 0     & 0     & 0     \\
                & 2.5   & D     & D     & D     & D     & D     & D     & D     & D     \\
                &       & 0.034 & 0.43  & 0.034 & 0.46  & 0.035 & 0.55  & 0.034 & 0.58  \\
                &       & 0     & 0     & 0     & 0     & 0     & 0     & 0     & 0     \\
                & 3.5   & D     & C     & D     & C     & D     & D     & D     & D     \\
                &       & 2.5   & 31    & 2.9   & 52    & 3.1   & 55    & 3.2   & 70    \\
                &       & 0     & 0.07  & 0     & 0.02  & 0     & 0     & 0     & 0     \\
\multicolumn{10}{c}{}\\
%
%                          CW      MC      CW     MC      CW       MC     CW       MC
%--------------------------------------------------------------------------------------
3 ............  & 1.5   & D     & D     & D     & D     & D     & D     & D     & D     \\
                &       & 0.014 & 0.031 & 0.014 & 0.032 & 0.014 & 0.033 & 0.014 & 0.036 \\
                &       & 0     & 0     & 0     & 0     & 0     & 0     & 0     & 0     \\
                & 2.5   & D     & D     & D     & D     & D     & D     & D     & D     \\
                &       & 0.28  & 3.6   & 0.29  & 5.1   & 0.29  & 5.8    & 0.29  & 6.5    \\
                &       & 0     & 0     & 0     & 0     & 0     & 0     & 0     & 0     \\
                & 3.5   & C     & C     & C     & C     & D     & C     & D     & C     \\
                &       & 21.5  & 33    & 44.4  & 83    & 42.3  & 130   & 43.5  & 350   \\
                &       & 0.078 & 0.25  & 0.035 & 0.22  & 0     & 0.20  & 0     & 0.18  \\
\multicolumn{10}{c}{}\\
%
%                          CW      MC      CW     MC      CW       MC     CW       MC
%--------------------------------------------------------------------------------------
7 ............  & 1.5   & D     & C     & D     & C     & D     & C     & D     & C     \\
                &       & 1.0   & 2.9   & 3.0   & 6.6   & 4.2   & 10    & 5.9   & 21    \\
                &       & 0     & 0.02  & 0     & 0.02  & 0     & 0.02  & 0     & 0.02  \\
                & 2.5   & C     & C     & C     & C     & C     & C     & C     & C     \\
                &       & 9.6   & 6.3   & 22.5  & 10.5  & 35.5  & 21    & 83.1  & 60    \\
                &       & 0.26  & 0.50  & 0.26  & 0.47  & 0.26  & 0.47  & 0.25  & 0.41  \\
                & 3.5   & C     & C     & C     & C     & C     & C     & C     & C     \\
                &       & 10.5  & 6.0   & 31.1  & 22    & 51.3  & 38    & 131.3 & 80    \\
                &       & 0.57  & 0.78  & 0.51  & 0.70  & 0.48  & 0.67  & 0.49  & 0.67  \\
%\multicolumn{10}{c}{}\\
%\hline
\enddata
\tablenotetext{a}{
The results of Chernoff \& Weinberg (1990, CW) are taken
from their Table~5. MC denotes our Monte Carlo results.
The first line describes the final state of the cluster at
the end of the simulation: C indicates core collapse, while
D indicates disruption.
The second line gives the time to core collapse or disruption,
in units of $10^9$ yr.
The third line gives the final cluster mass
in units of the initial mass.}
\end{deluxetable}

\begin{deluxetable}{lccc}
\tablecaption{Comparison of disruption times for
infinite ($N \to \infty$) and finite ($N = 3\times 10^5$)
F-P models from TPZ with Monte Carlo ($N = 3\times 10^5$) models.$^{\rm a}$}
\tablewidth{0pt}
\tablehead{
\colhead{  } & \colhead{\FP} & \colhead{\FP}
& \colhead{Monte Carlo} \\
\colhead{   } & \colhead{($N \to \infty$)} & \colhead{($N = 3\times 10^5$)}
& \colhead{($N = 3\times 10^5$)}
}
\startdata
$W_0 = 1$, Family~1     & $3.1 \times 10^{8}$ yr        & $4.8 \times 10^{8}$ yr        & $4.3 \times 10^{8}$ yr        \\
$W_0 = 1$, Family~4     & $3.3 \times 10^{8}$ yr        & $12.2 \times 10^{8}$ yr       & $5.8 \times 10^{8}$ yr        \\
\\
$W_0 = 3$, Family~1     & $2.2 \times 10^{9}$ yr        & $2.6 \times 10^{9}$ yr        & $3.6 \times 10^{9}$ yr        \\
$W_0 = 3$, Family~4     & $3.1 \times 10^{9}$ yr        & $5.3 \times 10^{9}$ yr        & $6.5 \times 10^{9}$ yr       \\
\enddata
\tablenotetext{a}{All models have a mass function $m^{-\alpha}$ with $\alpha = 2.5$
($\bar m = 1 \msun$).}
\end{deluxetable}

% -----------------------------  Figures  --------------------- %

\clearpage
\begin{figure}[t]
\plotone{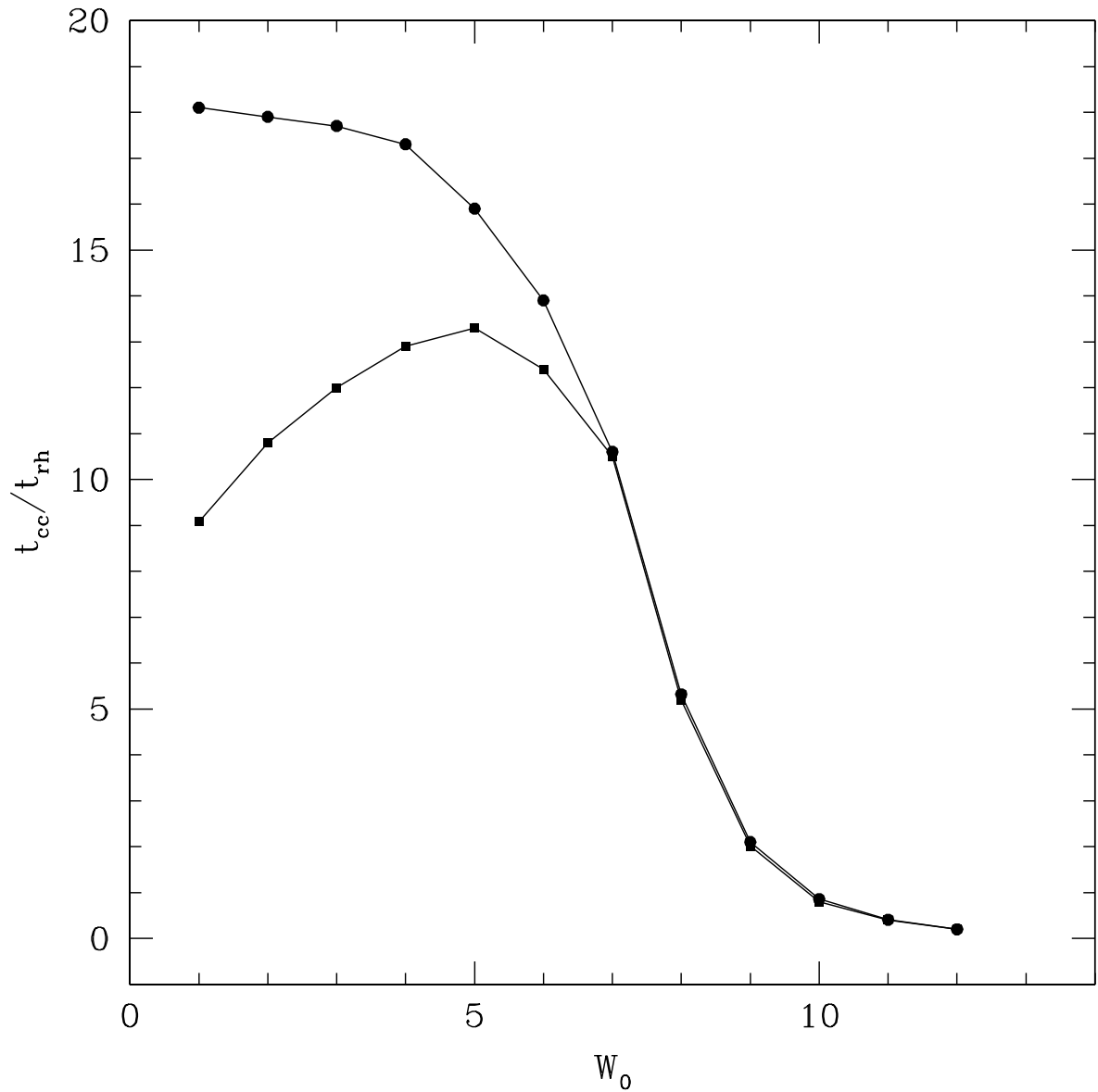}
\caption{Comparison of core-collapse times for $W_0 = 1-12$
single-component King models. Isolated models,  i.e., without an
enforced tidal boundary, are indicated by solid circles, while
tidally truncated models are indicated by squares.
\label{fig1}}
\end{figure}

\clearpage
\begin{figure}[t]
\plotone{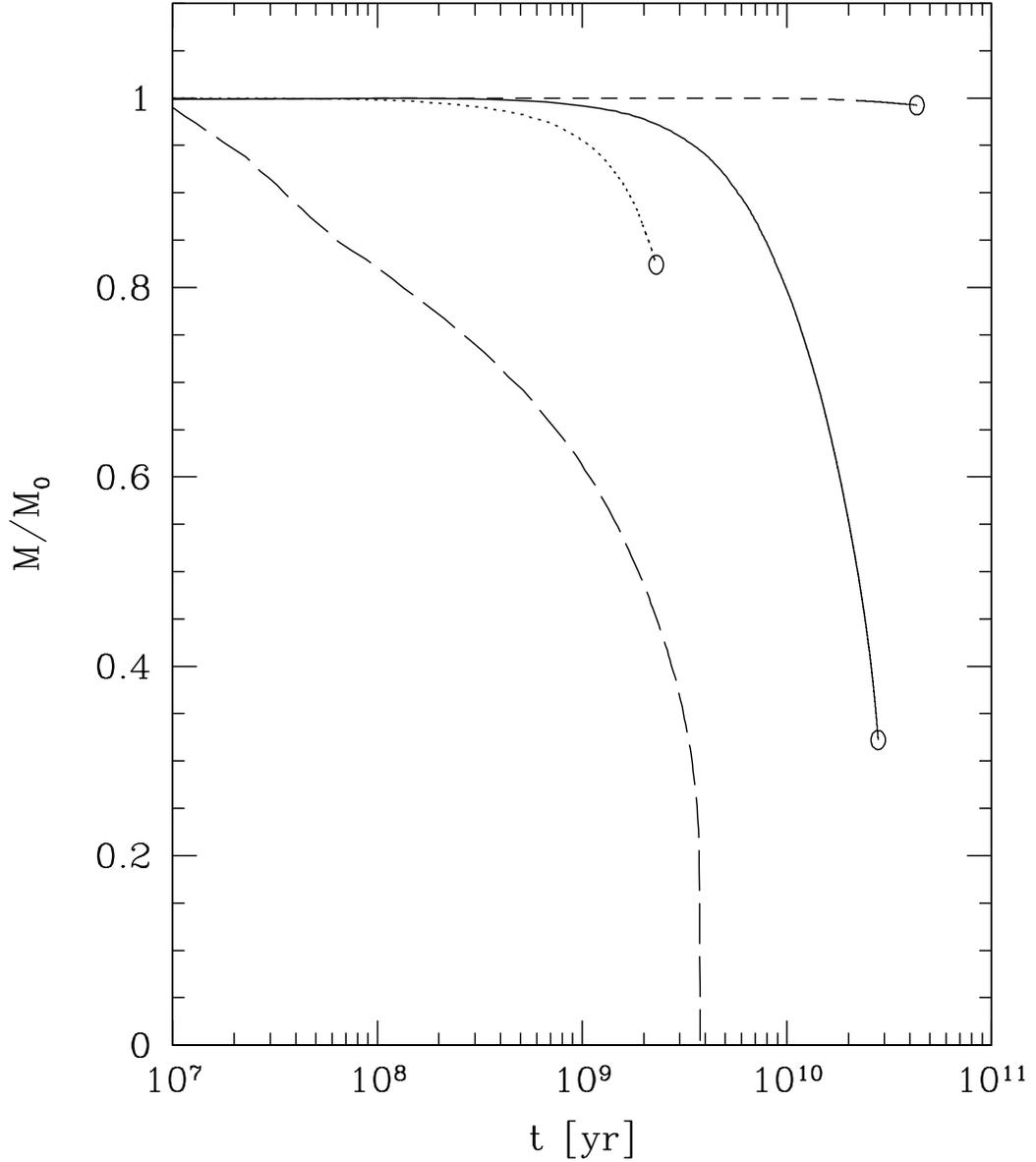}
\caption{Comparison of the mass loss rate in a $W_0 = 3$ King model due
to a tidal boundary, a power-law mass spectrum, and stellar evolution.
The mass of the cluster, in units of the initial mass $M_0$, is shown
as a function of time.
The solid and short-dashed lines are for a single-component model, with and
without a tidal boundary (Family~1), respectively.
The dotted line shows a model with a power-law mass spectrum, with
$\alpha = 2.5$, and a tidal boundary. The long-dashed line is for a more
realistic model with a tidal boundary, power-law mass spectrum, and stellar
evolution. The circle at the end of the line indicates core collapse. The
line without a circle indicates disruption of the cluster.
\label{fig2}}
\end{figure}

\clearpage
\begin{figure}[t]
\plotone{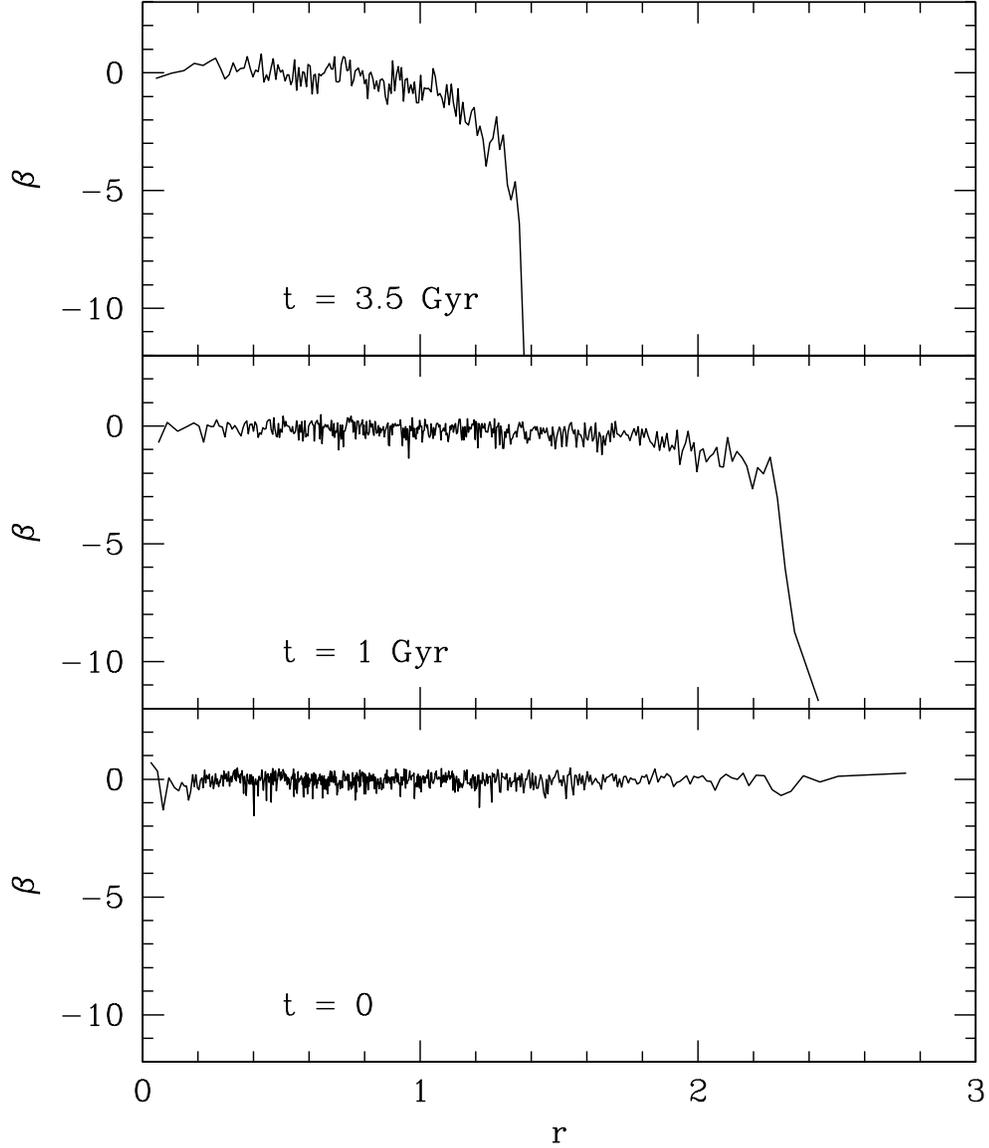}
\caption{Evolution of the anisotropy parameter
$\beta = 1 - \sigma_t^2/\sigma_r^2$ for a $W_0 = 3$ King model
($\alpha = 2.5$, Family~1).
The bottom frame shows the initial isotropic King model.
The top frame shows the anisotropy just before disruption.
The radius is in units of the virial radius. Stars on highly eccentric
orbits with large
apocenter distances in the cluster are preferentially removed, causing
$\sigma_t^2/\sigma_r^2$ to increase in the outer region.
\label{fig3}}
\end{figure}

\clearpage
\begin{figure}[t]
\plotone{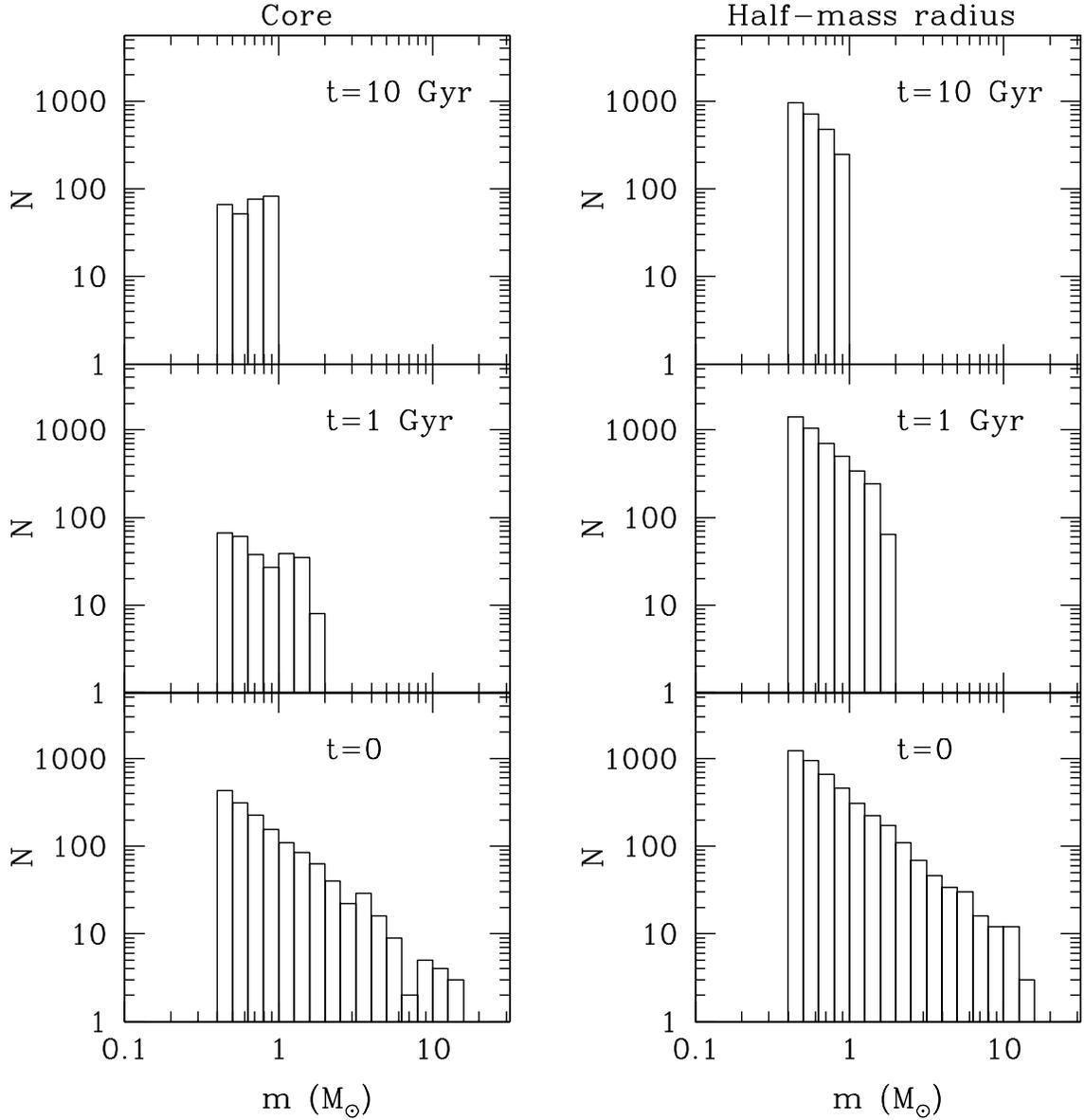}
\caption{Evolution of the main-sequence mass spectrum for a $W_0 = 7$ King model
with an initial
power-law mass function $m^{-\alpha}$, with $\alpha = 2.5$, Family~2.
The mass spectra in the core (left panels), and at the half-mass radius
(right panels) are shown at 1 Gyr, and 10 Gyr. The mass spectrum in the core
flattens dramatically as a result of stellar evolution, mass segregation
and evaporation.
\label{fig4}}
\end{figure}

\clearpage
\begin{figure}[t]
\plotone{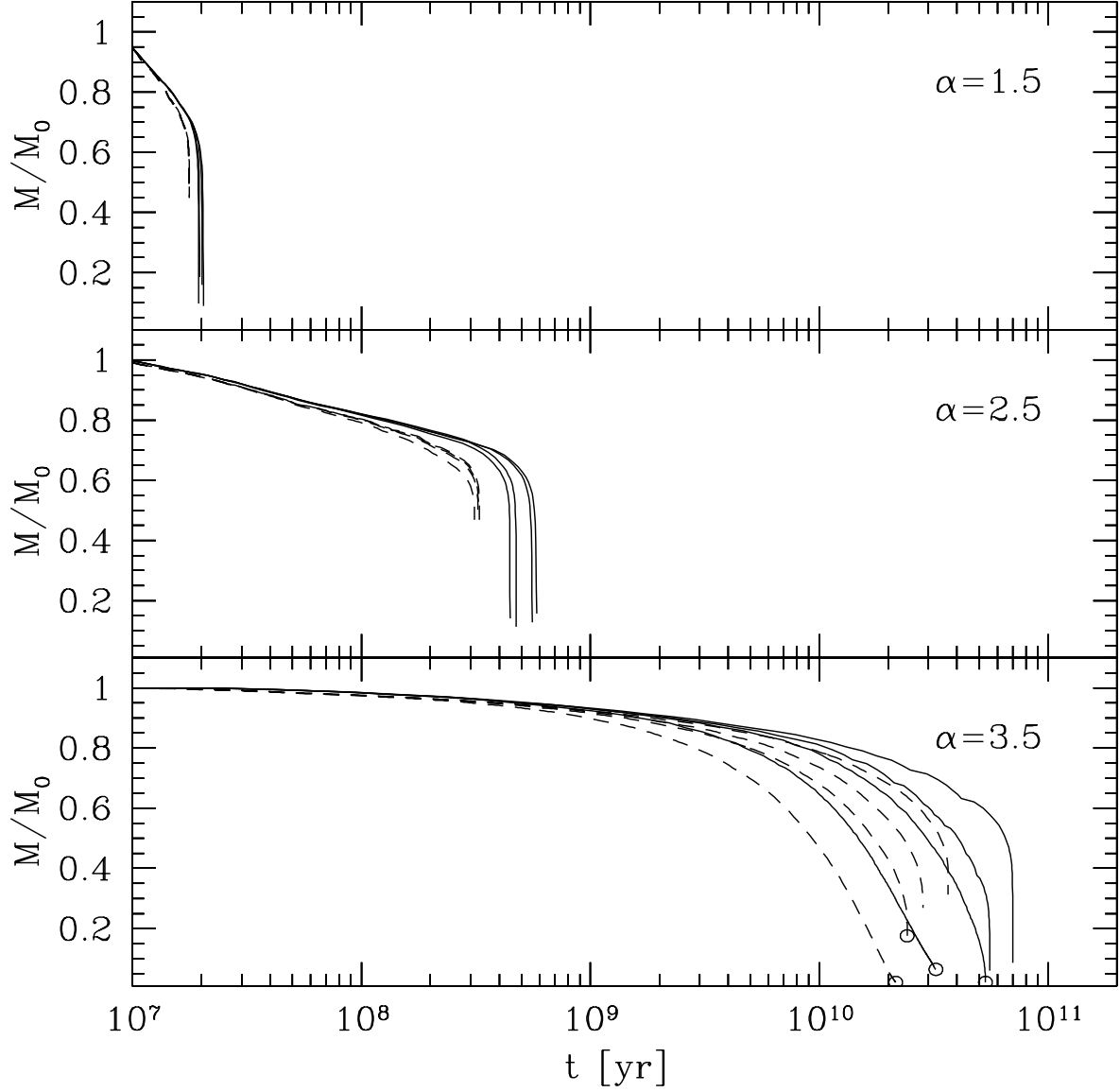}
\caption{Evolution of the total mass with time for $W_0 = 1$  King models,
Families~1--4. Comparison is made between our Monte Carlo models (solid lines)
and 2-D F-P models (dashed lines).
The three panels show results for different values of the exponent
$\alpha$ of the \emph{initial} power-law mass function ($m^{- \alpha}$).
The four lines for each case, represent Families~$1-4$, from left to right.
We indicate a core collapsed model with a circle at the
end of the line. Lines without a circle at the end indicate disruption.
\label{fig5}}
\end{figure}

\clearpage
\begin{figure}[t]
\plotone{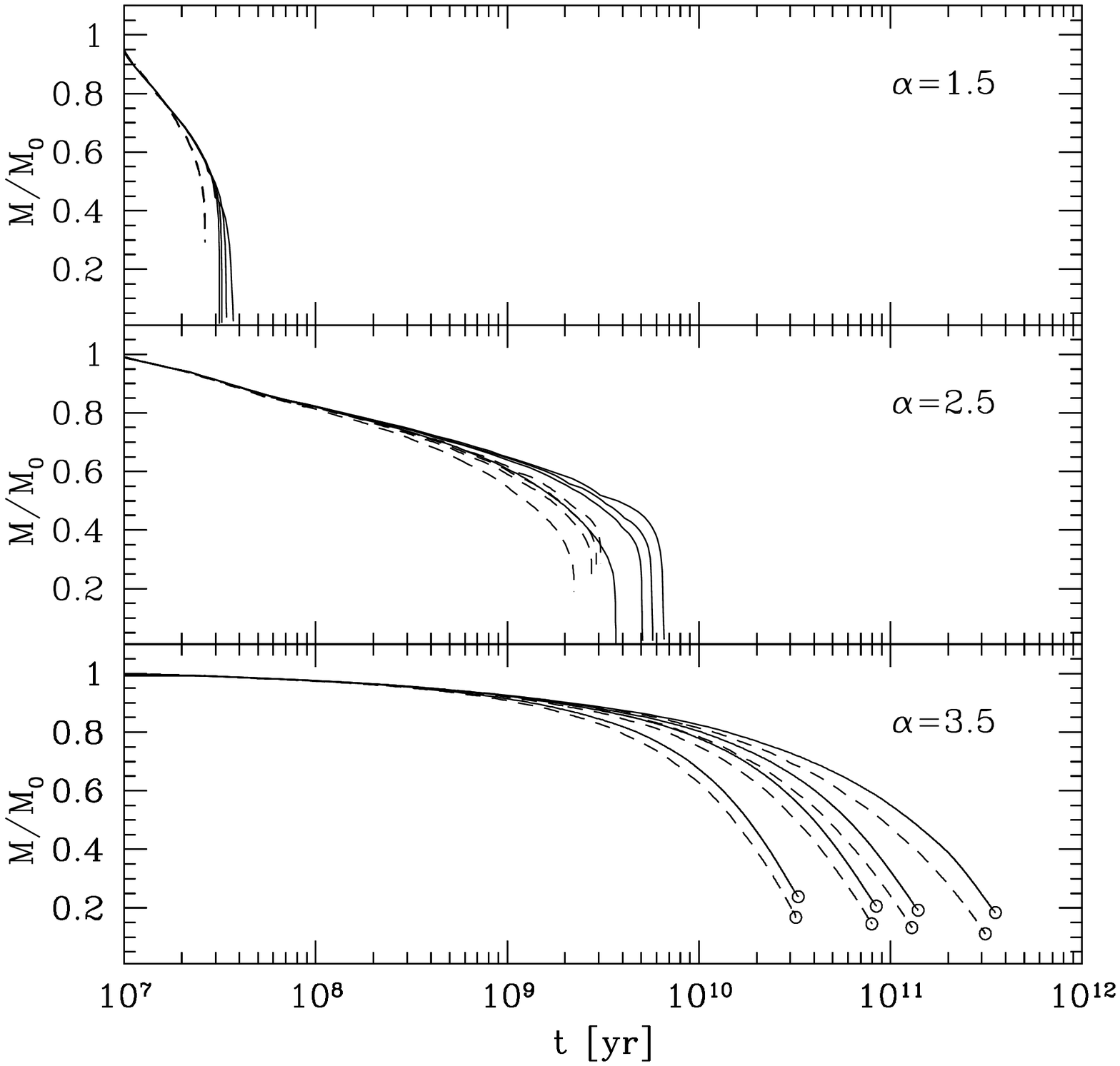}
\caption{Same as Figure~5, but for $W_0 = 3$ King models.
\label{fig6}}
\end{figure}

\clearpage
\begin{figure}[t]
\plotone{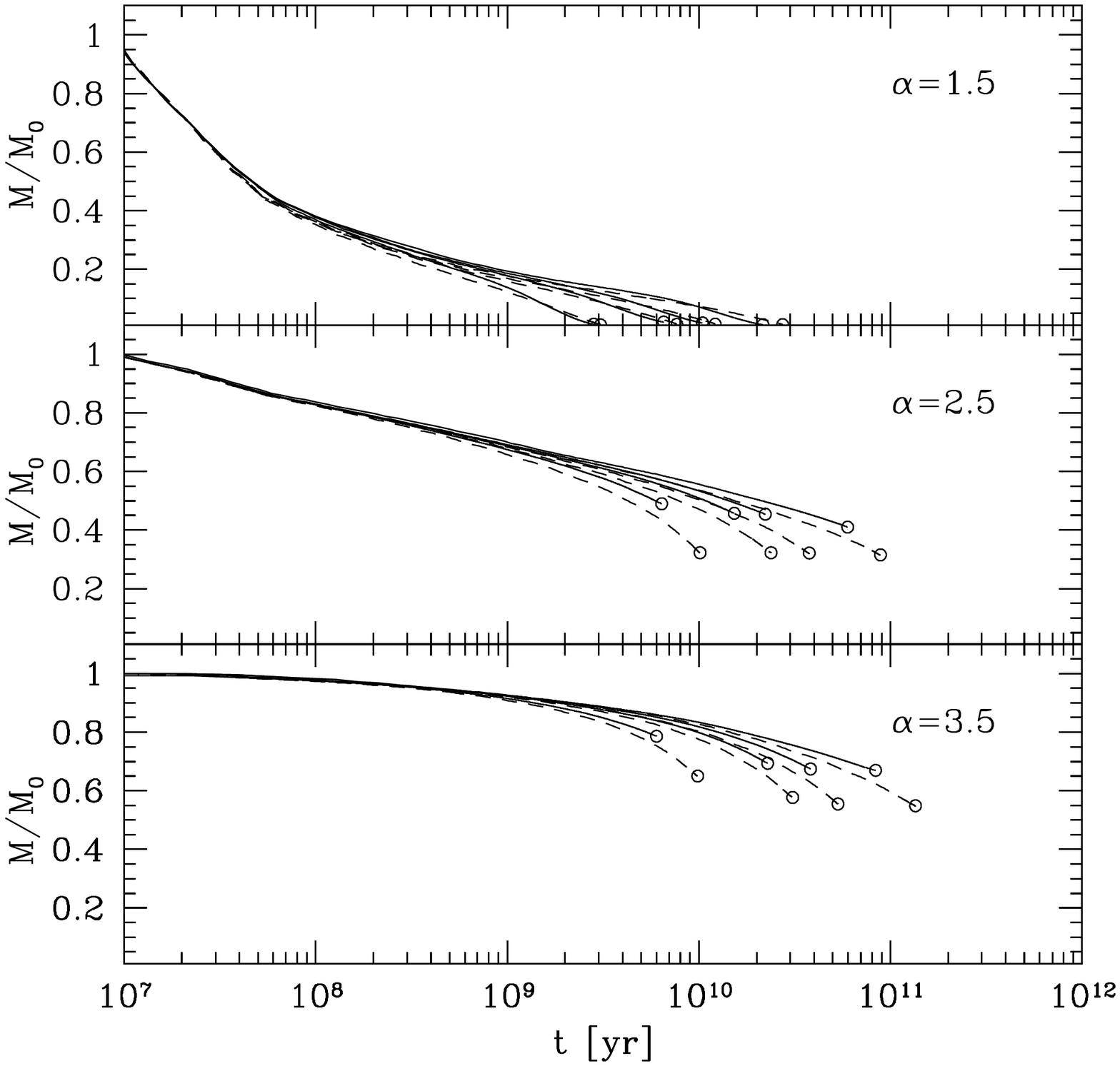}
\caption{Same as Figure~5, but for $W_0 = 7$ King models.
\label{fig7}}
\end{figure}

\clearpage
\begin{figure}[t]
\plotone{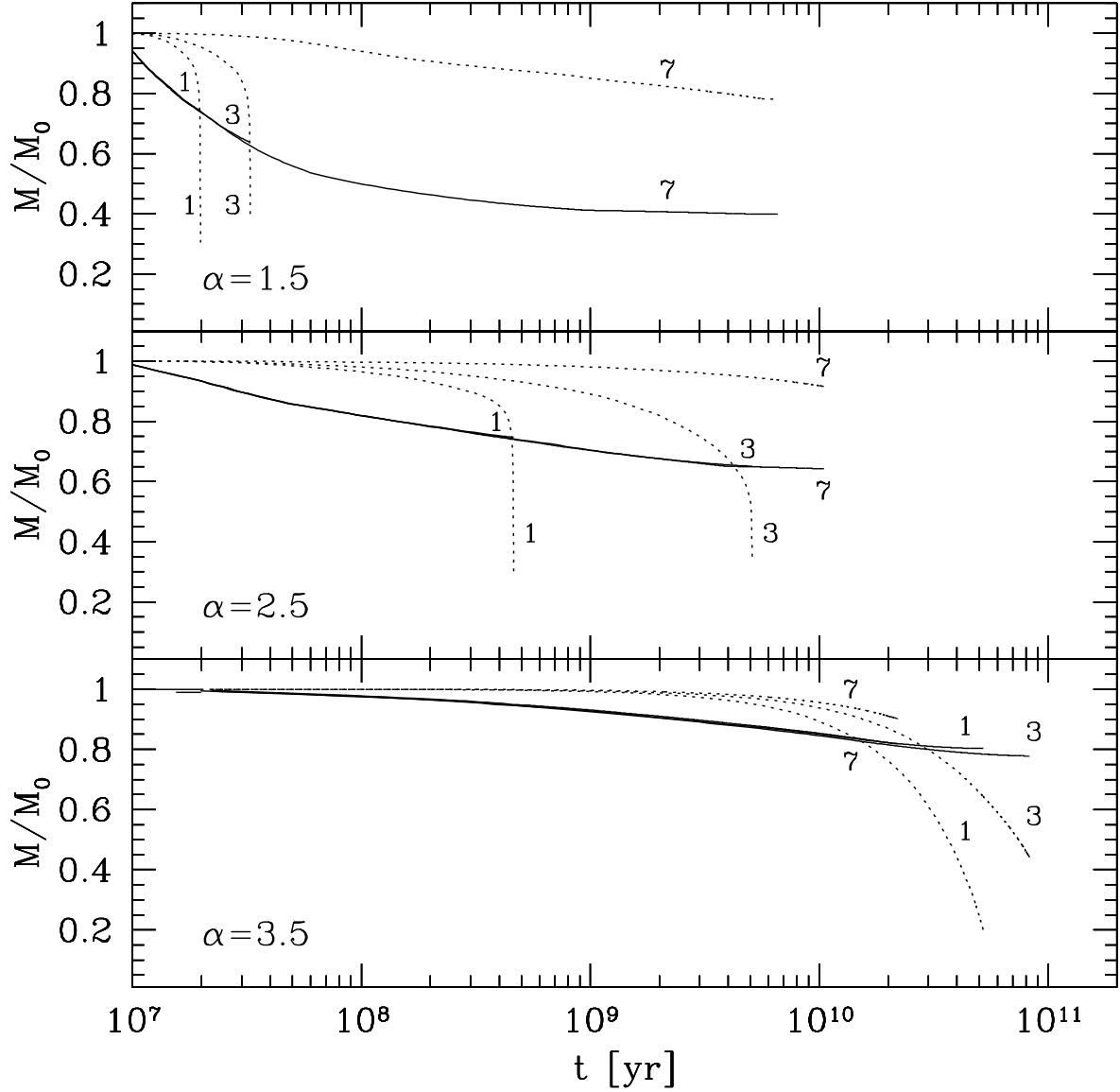}
\caption{Comparison of the mass loss due to stellar evolution (solid lines),
and mass loss due to tidal stripping of stars (dotted lines), for $W_0 = $
1, 3, and 7 King models,
with initial mass functions $m^{-\alpha}$, $\alpha =$ 1.5, 2.5, and 3.5.
The numbers 1, 3, and 7 next to the lines indicate an initial
model with $W_0 = $1, 3, and 7, respectively.
All models belong to Family~2. Results for other Families show similar trends.
Note that the mass loss due to stellar evolution is almost independent of
$W_0$ (as expected), but the
tidal mass loss varies significantly with $W_0$.
In the early phases of evolution, the mass loss due to stellar evolution
dominates, while in the later stages, tidal stripping of stars is the
dominant mechanism.
\label{fig8}}
\end{figure}

\clearpage
\begin{figure}[t]
\plotone{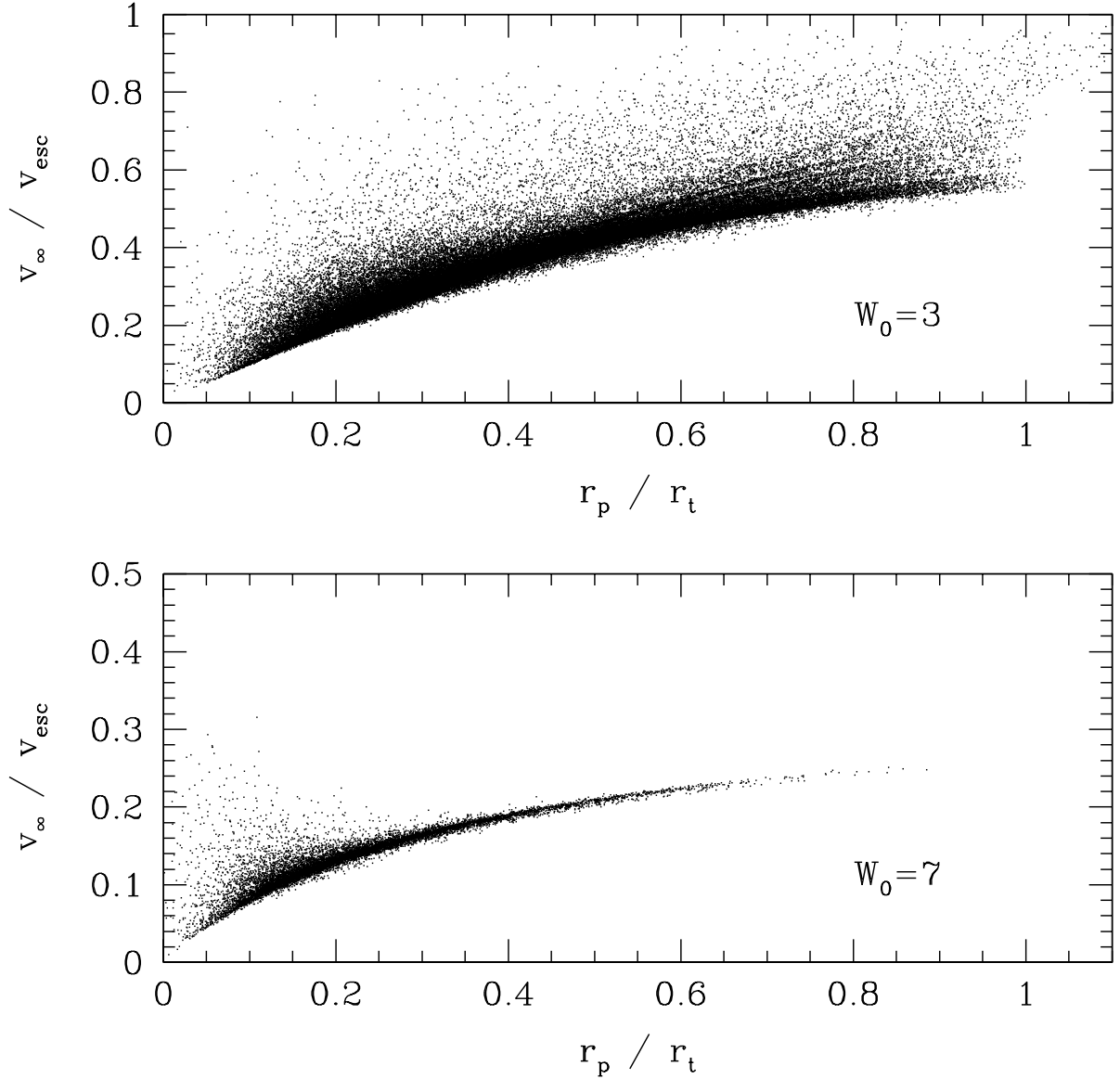}
\caption{Distribution of the pericenter distance and velocity of the escaping
stars, for two different King models: $W_0 = 3$ and 7 (Family~1, $\alpha = 2.5$).
The $W_0 = 3$ model (top frame) disrupts, while the $W_0 = 7$ model (bottom frame)
undergoes core collapse. The pericenter distance is given in units of
the initial tidal radius of the cluster. The velocity ``at infinity'' is
computed as $v_{\infty} = \sqrt{2(E-\phi_t)}$,
where $E$ is the energy per unit mass of the star, and $\phi_t$ is the
potential at the tidal radius. The escape velocity is defined as
$v_{esc} = \sqrt{2(\phi_t - \phi_0)}$, where $\phi_0$ is the potential at
the center of the cluster. The distribution of escape
velocities looks significantly different in the two clusters.
In the disrupting cluster ($W_0 = 3$),
the escaping stars have a wide range of escape energies at all pericenter
distances, whereas in the
collapsing cluster ($W_0 = 7$), a large fraction of the stars escape with
close to the minimum energy. Only the escapers from within the
central region have a significant range of escape energies.
\label{fig9}}
\end{figure}

\clearpage
\begin{figure}[t]
\plotone{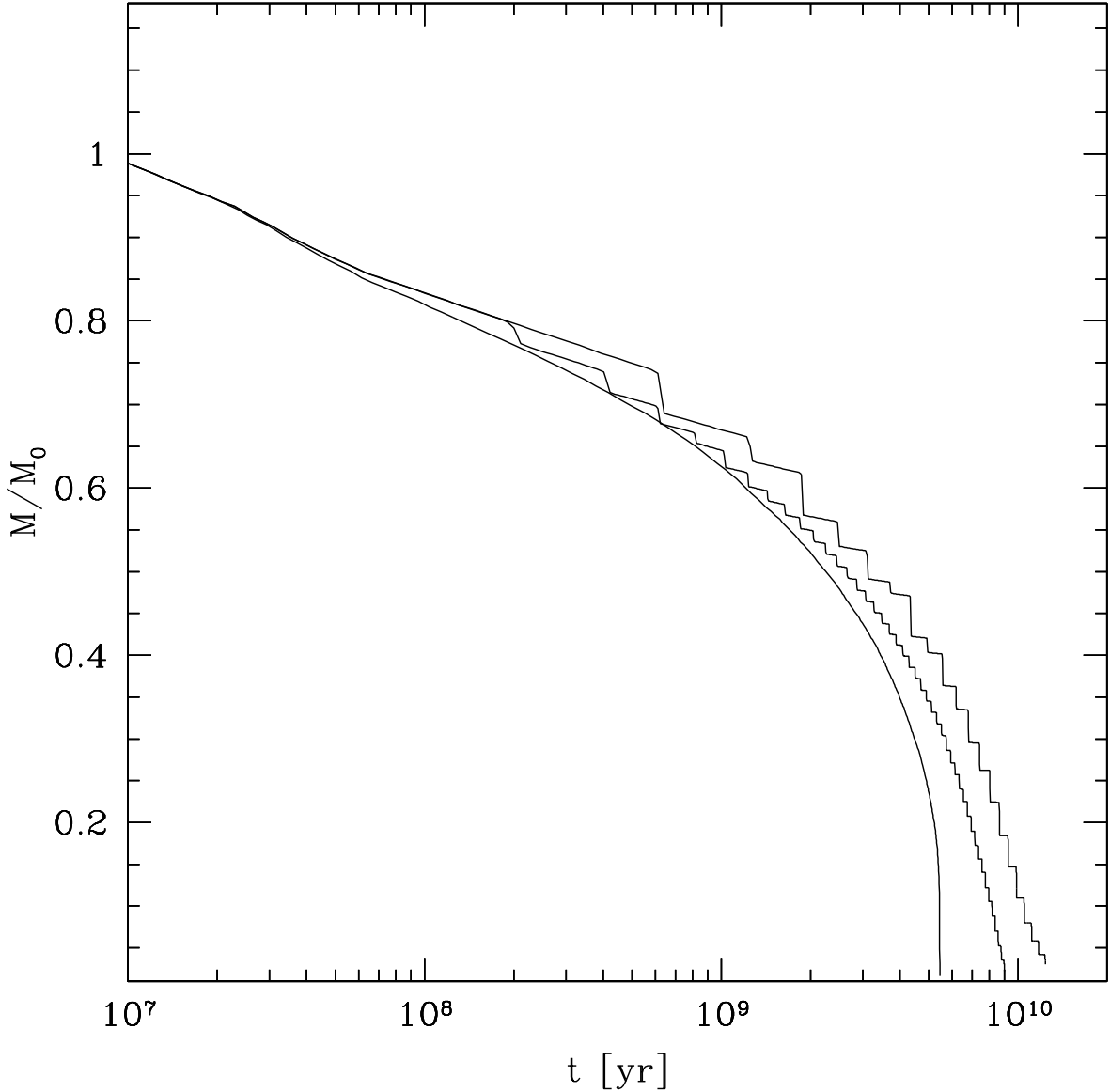}
\caption{Comparison of the mass loss for a $W_0 = 3$, $\alpha = 2.5$ (Family~2)
King model, on three different assumed orbits in the Galaxy. The leftmost
line shows a circular orbit, with radius $R_g = 5.76$ kpc. The cluster is
assumed to fill its Roche lobe at this distance. The rightmost line shows
a Keplerian elliptical orbit with eccentricity 0.6 and a \emph{pericenter}
distance of 5.76 kpc. Since the cluster on such an orbit spends most of its time
at a larger distance, the cluster does not fill its Roche lobe at all times.
This results in a sharp mass loss every time the cluster approaches pericenter.
The lifetime of the cluster is longer by almost a factor of 2. The intermediate
line is for an orbit in a more realistic Galactic potential, with a constant
circular velocity of $220 \kms$, with the same pericenter and apocenter distances
as for the Keplerian elliptical orbit. The orbit is no longer elliptical, and the
orbital period is shorter, resulting in a lifetime that is intermediate between
the circular and elliptical cases.
\label{fig10}}
\end{figure}


\begin{thebibliography}{}

\bibitem[Aarseth 1999]{a99}
Aarseth, S.~J. 1999, PASP, 111, 1333

\bibitem[Binney \& Tremaine 1987]{bt87}
Binney, J., \& Tremaine, S. 1987, Galactic Dynamics (Princeton, PUP)

\bibitem[Chernoff \& Weinberg 1990]{cw90}
Chernoff, D.~F. \& Weinberg, M.~D. 1990, ApJ, 351, 121

\bibitem[Cohn 1979]{c79}
Cohn, H. 1979, ApJ, 234, 1036

\bibitem[Cohn 1980]{c80}
Cohn, H. 1980, ApJ, 242, 765

\bibitem[Drukier 1995]{d95}
Drukier, G.~A. 1995, ApJS, 100, 347

\bibitem[Drukier \etal 1999]{dcly99}
Drukier, G.~A., Cohn, H.~N., Lugger, P.~M., \& Yong, H. 1999, ApJ, 518, 233

\bibitem[Fukushige \& Heggie 1995]{fh95}
Fukushige, T., \& Heggie, D.~C. 1995, MNRAS, 276, 206

\bibitem[Gao \etal 1991]{ggcm91}
Gao, B., Goodman, J., Cohn, H., \& Murphy, B. 1991, ApJ, 370, 567

\bibitem[Gnedin, Lee \& Ostriker 1999]{glo99}         % tidal shocking of clusters
Gnedin, O.~Y., Lee, H.~M., \& Ostriker, J.~P. 1999, ApJ, 522, 935

\bibitem[Goodman \& Hut 1989]{gh89}	% GC evolution supported by primordial binaries (semi-analytic estimates)
Goodman, J., \& Hut, P. 1989, Nature, 339, 40

\bibitem[Heggie \etal 1999]{hgst99}     % collaborative experiment
Heggie, D.~C., Giersz, M., Spurzem, R., \& Takahashi, K. 1999, in Highlights
of Astronomy, vol. 11, ed. J.~Andersen, 591

\bibitem[Henon 1971a]{h71a}
Henon, M. 1971a, Ap. Space Sci., 13, 284

\bibitem[Henon 1971b]{h71b}
Henon, M. 1971b, Ap. Space Sci., 14, 151


\bibitem[Hut, McMillan, \& Romani 1992]{hmr92}
Hut, P., McMillan, S., \& Romani, R.~W. 1992, ApJ, 389, 527


\bibitem[Iben \& Renzini 1983]{ir83}
Iben, I., \& Renzini, A. 1983, Ann. Rev. Astr. Ap., 21, 271


\bibitem[Joshi \etal 1999]{j99}
Joshi, K.~J., Rasio, F.~A., \& Portegies Zwart, S. 2000, ApJ, 540, 969 [Paper I]


\bibitem[Lee \& Ostriker 1987]{lo87}
Lee, H.~M., \& Ostriker, J.~P. 1987, ApJ, 322, 123


\bibitem[Makino \etal 1997]{m97}
Makino, J., Taiji, M., Ebisuzaki, T., \& Sugimoto, D. 1997, ApJ, 480, 432

\bibitem[McMillan \& Hut 1994]{mh94}    % GC evolution with primordial binaries - III - galactic tidal field
McMillan S.~L.~W., \& Hut P. 1994, ApJ, 427, 793


\bibitem[McMillan, Makino, \& Hut 1990]{mmh90}    % GC evolution with primordial binaries - I
McMillan S.~L.~W., Hut P., \& Makino, J. 1990, ApJ, 362, 522


\bibitem[Meylan \& Heggie 1997]{mh97}
Meylan, G., \& Heggie, D.~C. 1997, A\&A Rev. 8, 1


\bibitem[Miller \& Scalo 1979]{ms79}
Miller, G.~E., \& Scalo, J.~M. 1979, ApJ, 41, 513


\bibitem[Odenkirchen \etal 1997]{obgt97}	% GC orbits -- typical eccentricity is 0.6 (~0.2 -- 0.97)
Odenkirchen, M., Brosche, P., Geffert, M., \& Tucholke, H.~J. 1997, New Astronomy, 2, 477

\bibitem[Portegies Zwart \etal 1998]{pzhmm98}
Portegies Zwart, S., Hut, P., Makino, J., \& McMillan, S.L.W. 1998, A\&A, 337, 363


\bibitem[Quinlan 1996]{q96}
Quinlan, G.D. 1996, New Astronomy, 1, 255

\bibitem[Ross \etal 1997]{rmh97}
Ross, D.~J., \& Mennim, A., \& Heggie, D.~C. 1997, MNRAS, 284, 811


\bibitem[Spitzer 1987]{s87}
Spitzer, L. 1987, Dynamical Evolution of Globular Clusters
(Princeton: Princeton University Press)

\bibitem[Spitzer \& Mathieu 1980]{sm80} 	% GC evolution with primordial binaries
Spitzer, L., \& Mathieu, R.~D. 1980, ApJ, 241, 618


\bibitem[Takahashi 1995]{t95}
Takahashi, K. 1995, PASJ, 47, 561


\bibitem[Takahashi 1996]{t96}
Takahashi, K. 1996, PASJ, 48, 691


\bibitem[Takahashi 1997]{t97}
Takahashi, K. 1997, PASJ, 49, 547


\bibitem[Takahashi \& Portegies Zwart 1998]{tpz98}
Takahashi, K., \& Portegies Zwart, S.~F. 1998, ApJ, 503, L49


\bibitem[Takahashi \& Portegies Zwart 1999]{tpz99}
Takahashi, K., \& Portegies Zwart, S.~F. 2000, ApJ, 535, 759 [TPZ]


\end{thebibliography}
\end{document}